# Right Device, Right Place: Variation-Aware Placement and Design Optimization for Robust Photonic Lattice-Filter Demultiplexers

Zahra Ghanaatian, Edwin K. P. Chong, and Mahdi Nikdast
Department of Electrical and Computer Engineering, Colorado State University, Fort Collins, CO 80523 USA
mahdi.nikdast@colostate.edu

***Abstract:*** *We propose a Bayesian co-optimization framework for robust integrated photonic lattice-filter demultiplexers, jointly optimizing device placement and design parameters under fabrication and thermal variations. Results show 75% better spectral matching and 45% lower calibration power.*

## I. Introduction

As AI-era communication networks increasingly rely on wavelength-parallel optical links to move data across packages, boards, and datacenter fabrics, integrated photonic demultiplexers and interleavers are becoming critical building blocks for channel separation, spectral routing, bandwidth aggregation, and scalable Wavelength-Division Multiplexing (WDM) resource management. One widely used approach is to cascade Mach-Zehnder Interferometer (MZI) lattice filters to synthesize flat-top passbands that improve channel isolation and usable spectral bandwidth [1], [2]. However, these devices are sensitive to thermal and fabrication-process variations (FPVs), resulting in channel distortion, crosstalk noise, and increased calibration power. Furthermore, in large-scale circuits with cascaded devices, such as MZI-based demultiplexers, inter-device spectral alignment is critical but challenging due to the nonuniformity of variations that devices experience, leading to even more spectral errors and calibration complexity and power.

Prior work has attempted to mitigate FPVs and thermal variations by optimizing waveguide dimensions or utilizing silicon nitride (SiN) layers [3–6]. However, these approaches fail to handle simultaneous FPVs and thermal variations, nor do they resolve inter-device spectral alignment. To address this, we propose a novel Random Forest-based Bayesian Optimization (BO) framework for an O-band 1×8 flat-passband lattice-filter-based demultiplexer, considered as a case-study in this paper. The framework jointly optimizes device placement and design parameters under FPVs and thermal variations to maximize inter-device spectral alignment. To the best of our knowledge, this is the first framework to perform variation-aware placement and design optimization in Photonic Integrated Circuits (PICs). Our results show 75% improvement in inter-device spectral uniformity and 45% saving in calibration power. Note that our framework can be extended to other PICs, and the demultiplexer is considered as a case-study in this paper.

## II. Variation-Aware Robust Design and Placement Co-Optimization

A schematic layout of the eight-channel flat-passband lattice-filter-based demultiplexer with a channel spacing of 4.4 nm is shown in Fig. 1(a). The circuit includes seven MZI-based wavelength splitters (11 MZIs in total), cascaded to form

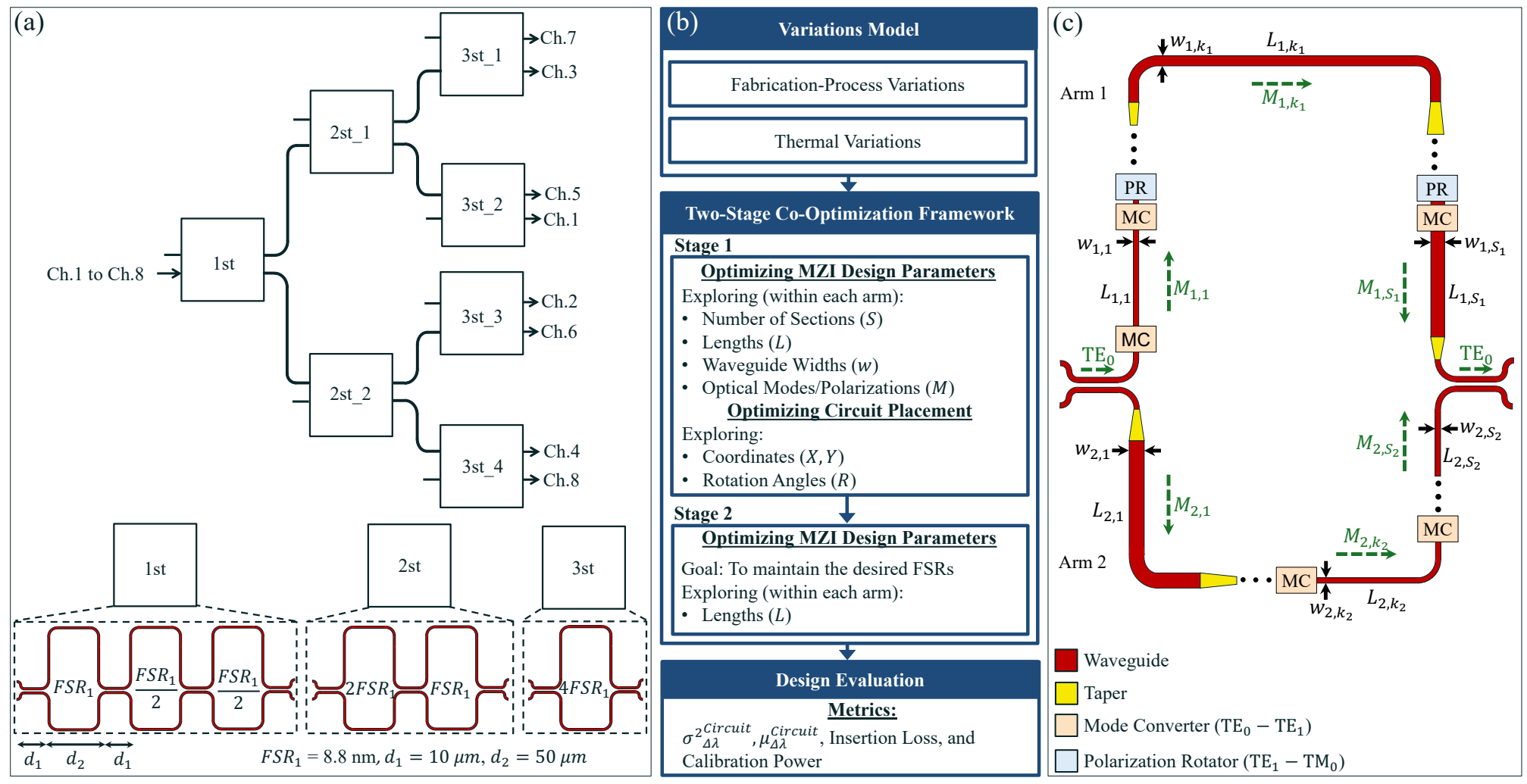

Fig. 1. (a) Schematic layout of the eight-channel flat-passband lattice-filter-based demultiplexer. Each square represents an MZI-based wavelength splitter with one, two, or three delay stages, as indicated by the labels in the squares. (b) Overview of our proposed Random Forest-based Bayesian Optimization framework. (c) Example schematic layout and design parameters of an optimized MZI. PR: Polarization rotator. MC: Mode converter.

This work was supported in part by the NSF under grant numbers CNS-2046226 and ECCS-2520399.

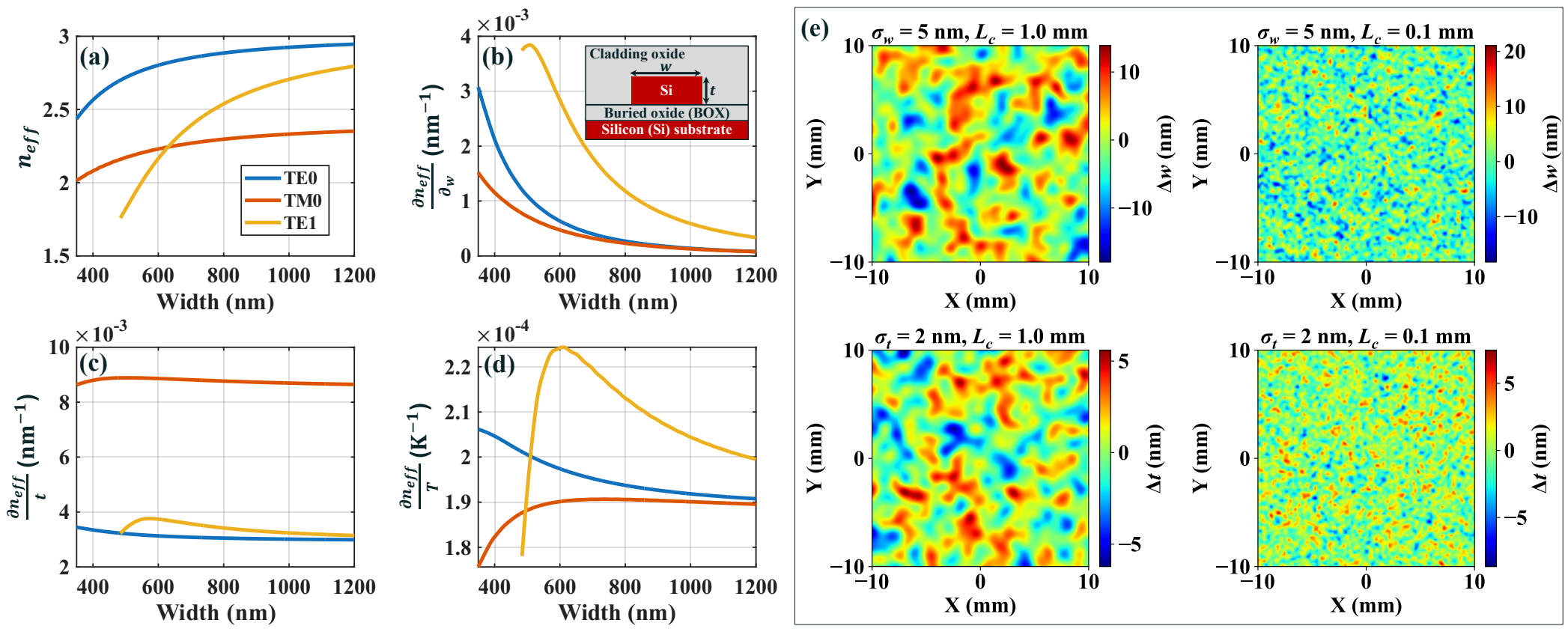

Fig. 2. (a) Effective index ($n_{eff}$) of a silicon-on-insulator (SOI) strip waveguide as a function of waveguide width for the TE0, TE1, and TM0 modes at λ = 1310 nm and silicon thickness of $t$ = 220 nm. Sensitivity of the effective index to variations in (b) waveguide width ($w$) with inset showing the waveguide cross-section, (c) thickness ($t$), and (d) temperature ($T$), plotted versus waveguide width range of 350 nm to 1200 nm (considered as an example here). (e) Digital-twin variation maps for $\sigma_w$ = 5 nm, $\sigma_t$ = 2 nm, shown for correlation lengths of $L_c$ = 0.1 mm and $L_c$ = 1 mm.

a three-stage binary-tree structure. To achieve eight-channel (de)multiplexing, three filters with distinct Free Spectral Ranges (FSRs) are cascaded for sequential filtering. Each filter is designed with optimized coupling coefficients and delay-line lengths, while the FSR doubles at each successive stage order [1], [2].

To maximize inter-device spectral matching under FPVs and thermal variations, thereby improving demultiplexer robustness, we propose a variation-aware Bayesian-optimization-based co-optimization framework shown in Fig. 1(b). For the demultiplexer in Fig. 1(a), the framework jointly optimizes the design parameters of all the 11 MZIs, as illustrated in Fig. 1(c), together with the placement and rotation of the seven wavelength splitters. Note that the spacing between the MZIs within each wavelength splitter ($d_1$) is assumed to be fixed. The BO is performed as follows. An initial set of designs is sampled from the design space and evaluated to construct the initial dataset. A Random Forest surrogate model is then fitted to the dataset to approximate the objective function, and new designs are iteratively selected by maximizing the Expected Improvement (EI) acquisition function. Each selected design is evaluated by the simulator, and the resulting data are incorporated into the dataset to update the surrogate model. The process continues until the predefined simulation budget (e.g., number of iterations) is exhausted, after which the design yielding the minimum objective function value is identified as the optimum. The optimization problem at the heart of our framework is the following:

$$\min_{x} \sigma^{2}{}^{Circuit}_{\Delta\lambda}(x) + \Delta FSR\,(x), \qquad (1)$$
$$s.t.\ f(x) \leq B, g_{Overlap}(x) \geq g_{min}.$$

Here, the decision vector is defined as $x = [S, L, w, M, P, R]$, where $S$ denotes the number of sections in the MZI arms, $L$ represents the section lengths, $w$ corresponds to the waveguide widths, $M$ specifies the optical mode/polarization (TE0, TE1, and TM0), $P$ defines the $(X, Y)$ coordinates of wavelength splitters, and $R$ is the rotation angle of wavelength splitters. In addition, in (1), $f$ denotes the circuit footprint, constrained by a bounding box ($B$). The constraints $g_{Overlap}$ and $g_{min}$ enforce non-overlap conditions and minimum spacing between wavelength splitters, respectively. The objective minimizes the variance of central wavelength shift across all the MZIs ($\sigma^{2}{}^{Circuit}_{\Delta\lambda}$) and the worst-case FSR error ($\Delta FSR$). The wavelength shift ($\Delta\lambda$) and FSR are modeled as follows [7], [8]:

$$\Delta\lambda = \lambda \sum_{\substack{k_1=1\\ X=w,t,T}}^{S_1} \sum_{\substack{k_2=1\\ X=w,t,T}}^{S_2} \left| \frac{L_{1,k_1}\frac{\partial n_{eff_{1,k_1}}}{\partial X}\left|\rho_{X_{1,k_1}}\right| - L_{2,k_2}\frac{\partial n_{eff_{2,k_2}}}{\partial X}\left|\rho_{X_{2,k_2}}\right|}{n_{g_{1,k_1}}L_{1,k_1} - n_{g_{2k_2}}L_{2,k_2}} \right|. \qquad (2)$$

$$FSR = \sum_{k_1=1}^{S_1} \sum_{k_2=1}^{S_2} \left| \frac{\lambda^2}{n_{g_{1,k_1}}L_{1,k_1} - n_{g_{2k_2}}L_{2,k_2}} \right|. \qquad (3)$$

Here, $\lambda$ is the central wavelength; $n_g$ is the group index; $k_1$, $k_2$ and $S_1$, $S_2$ denote the section indices and total number of sections in arm1 and arm2, respectively; and $L_{1,k_1}$ and $L_{2,k_2}$are the length of the corresponding sections. The terms $\rho_X$ (with $X = w, t, T$) represent variations due to waveguide width, thickness, and temperature variations from room temperature. The sensitivity terms $\frac{n_{eff}}{\partial X}$ capture the corresponding changes in the effective index. Figs. 2(a)–(d) show these sensitivities for TE0, TE1, and TM0 modes. Using the mean and 3σ values for thickness and width variations provided by a foundry, we can model $\rho_X$ as a normal distribution. This enables generation of spatial digital-twin variation maps for width and thickness (see Fig. 2(e)) using the method in [9]. During each optimization iteration, Monte Carlo simulations are performed by placing the circuit at random locations on the variation maps while applying random thermal variations. Then, (2) is used to compute $\Delta\lambda$ for each MZI. For each Monte Carlo sample, the circuit-level variance ($\sigma^{2}{}^{Circuit}_{\Delta\lambda}$) is computed across all the 11 MZIs, and the final value of $\sigma^{2}{}^{Circuit}_{\Delta\lambda}$ used in (1) is obtained by averaging over all samples.

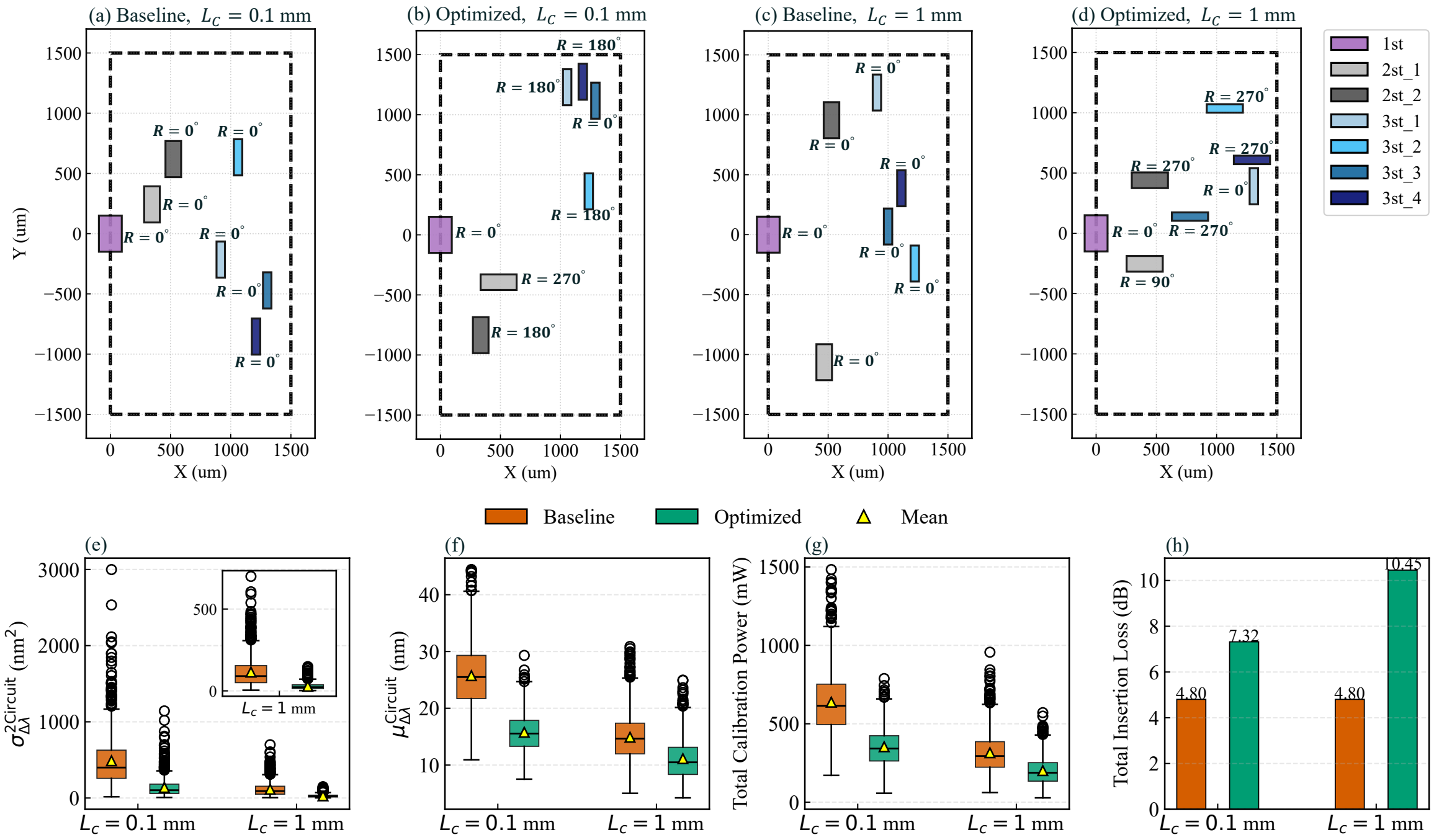

Fig. 3. Baseline and optimized placement of the wavelength splitters for (a–b) $L_c$ = 0.1 mm and (c–d) $L_c$ = 1 mm. The black dashed lines are bounding box $B$. (e) Circuit-level wavelength-shift variance ${\sigma^2}_{\Delta\lambda}^{Circuit}$, (f) mean $\mu_{\Delta\lambda}^{Circuit}$, (g) total calibration power, and (h) total insertion loss.

Although $\Delta FSR$ is minimized in (1), exact FSR matching may not be guaranteed for all the MZIs. Therefore, a second-stage post-optimization is performed (see Fig. 1(b)), where the optimized $x$ from stage one is fixed, and only section lengths are slightly adjusted to satisfy the target FSR. Finally, since the optimized design employs multiple waveguide widths and optical modes/polarizations, tapers, Mode Converters (MC), and Polarization Rotators (PR) are required between sections to ensure smooth transitions. All these devices can be inverse designed to guarantee low loss.

## III. Results and Discussion

The proposed framework is evaluated under the following design and simulation conditions. The maximum number of sections in each MZI is set to $S_1 = S_2 = 2$. The circuit footprint is constrained within a bounding box of $B = 1500 \times 3000\ \mu m^2$. A minimum spacing of $g_{min} = 15\ \mu m$ is enforced between adjacent wavelength splitters, and the rotation angles of wavelength splitters are restricted to $R \in \{0°, 90°, 180°, 270°\}$. FPVs are modeled using spatially correlated variations in waveguide width and thickness. Specifically, width variations follow a Gaussian distribution with standard deviation $\sigma_w = 5$ nm, and thickness variations follow $\sigma_t = 2$ nm. The standard deviations used in this study are derived from a confidential foundry process model and are not disclosed in this paper. Two correlation lengths, $L_c = 0.1$ mm (short) and $L_c = 1$ mm (long), are considered to capture different spatial variations. The resulting variation maps are shown in Fig. 2(e). Thermal variations are modeled independently for each MZI. Within a given MZI, both arms are assumed to experience identical thermal fluctuations randomly sampled from a uniform distribution in the range 0–80 $K$, *i.e.*, $\rho_{T_{1,k_1}} = \rho_{T_{2,k_2}} \sim U(0, 80\ K)$. Thermal variations are assumed to be uncorrelated across different MZIs. The optimization is performed over 700 iterations, with 1000 Monte Carlo samples per iteration. The baseline design uses 500-nm-wide waveguides and TE0 for all the MZIs. For the baseline, wavelength splitters are randomly placed within the bounding box with fixed orientation ($R = 0°$). The baseline and optimized circuit layouts are shown in Figs. 3(a)–(d) for both correlation lengths.

To evaluate robustness, a Monte Carlo analysis is conducted using 1000 random placements of each circuit instance across the FPV maps, combined with random thermal variations. For each instance, the wavelength shift ($\Delta\lambda$) of all the 11 MZIs is computed using (2), and the circuit-level statistics, including mean ($\mu$) and variance ($\sigma^2$), are extracted. The performance comparison between the baseline and optimized designs is summarized in Figs. 3(e)–(h), including ${\sigma^2}_{\Delta\lambda}^{Circuit}$, $\mu_{\Delta\lambda}^{Circuit}$, total calibration power (calculated based on the model in [10]), and total insertion loss. The optimized design achieves average improvements of 72%, 39%, and 45% in ${\sigma^2}_{\Delta\lambda}^{Circuit}$, $\mu_{\Delta\lambda}^{Circuit}$, and total calibration power, respectively, for $L_c = 0.1$ mm, and 75%, 25%, and 36% for $L_c = 1$ mm, compared to the baseline. Despite these improvements, the optimized circuit shows higher insertion loss due to additional components introduced: tapers, MCs, and PRs. Prior inverse-designed device implementations have demonstrated losses below 0.1 dB for the MC and PR [11], while taper loss is negligible when sufficiently long tapers are employed [7].

In summary, our results show that the proposed co-optimization framework significantly enhances inter-device matching and robustness against FPVs and thermal variations in integrated photonic demultiplexers. This facilitates post-fabrication tuning using either laser frequency detuning or a single global heater due to high inter-device matching.